\documentstyle[preprint,aps,latexsym,epsfig,graphicx,epsfig]{revtex}

\draft

\title{Concentration of higher dimensional entanglement}

\author{Alipasha Vaziri$^1$, Jian-Wei Pan$^1$, Thomas Jennewein$^1$, Gregor Weihs$^{1,2}$, and Anton Zeilinger$^1$}

\address{$^1$Institut f\"ur Experimentalphysik, Universit\"at Wien,
Boltzmanngasse 5, A--1090 Wien, Austria \\ $^2$Ginzton
Laboratory, S-23 Stanford University, Stanford, CA 94304, USA}

\date{\today}

\begin{document}

\maketitle

\begin{abstract}

Enhancement of entanglement enhancement is necessary for most
quantum communication protocols many of which are defined in
Hilbert spaces larger than two. In this work we present the
experimental realization of entanglement concentration of orbital
angular momentum entangled photons produced in the spontaneous
parametric down-conversion process which have been shown to
provide a source for higher dimensional entanglement. We
investigate the specific case of three dimensions and the
possibility of generating different entangled states out of an
initial state. The results presented here are of importance for
pure states as well as for mixed states.
\end{abstract}


Most of the current applications of entanglement in quantum
communication such as quantum teleportation \cite{Bennett93a} and
quantum cryptography\cite{Ekert91a,Gisin94a} work best for
maximally entangled states. However, in practice one always has
to deal with non maximally entangled or mixed states for example
because of the uncontrollable interaction of the particles with
the environment. Therefore many quantum protocols such as
distillation \cite{Horodecki97a,Kwiat01a} purification
\cite{Bennett96b}, concentration \cite{Bennett96a} and error
correction \cite{Shor95a,Stean96a} have been suggested in order
to enhance the quality of entanglement. With the exception of
quantum error correction which is most important for quantum
computation, the common idea in all other protocols is to start
with a sample of entangled states having a low quality of
entanglement and using local operations and classical
communication only to end up in a smaller ensemble with a higher
degree of entanglement. A measure for the enhancement of the
entanglement could for example be the non local feature of the
state represented by violation of Bell's inequalities.\par

There are varying terminologies in the literature. Following
reference \cite{Bennett96a} we will use the term concentration
for our experimental achievement which was to extract maximally
entangled states out of non maximally entangled pure states. This
method is also referred to as local filtering or the "Procrustean
method" \footnote{After Procrustes, a fabulous Greek giant who
stretched or shortened captives to fit one of his iron beds}. In
contrast to concentration the term distillation is used for the
more general case of extracting entanglement out of initially
mixed states. Purification denotes the procedure which makes
arbitrary initial states more pure, after local operations and
classical communication, but not necessarily more entangled. As
demonstrated by Horodecki et al. \cite{Horodecki97a}, even if the
fidelity of the system to be in the desired entangled state is
less than $1/2$, all inseparable quantum systems can be distilled
to singlet states by local filtering and the Bennett et al.
distillation protocol \cite{Bennett96b}. This means that the
Procrustean concentration method as experimentally demonstrated
here is not only of importance for pure states but also for
extracting entanglement out of mixed states.\par

There exists a steadily growing interest in entanglement in higher
dimensions since it allows realization of new types of quantum
communication protocols
\cite{Bartlett00a,Ambainis01a,Bechmann00a,Bourennane01a}. Such
states also have been shown to be more efficient and to provide
more security in quantum communication applications such as in
quantum cryptography \cite{Bourennane01a,Bechmann00a,Bechmann00b}.
However in order to overcome the problem of uncontrollable
interaction of the quantum states with the environment and to
make quantum communication with higher dimensional entangled
states feasible it is important to be in full control of the
corresponding distillation and concentration techniques. So far
the Procrustean method has been experimentally demonstrated only
for entangled qubits, i.e. for two-dimensional  systems
\cite{Kwiat01a}. Given the motivations discussed above we
demonstrate for the first time the experimental realization of
quantum concentration in higher dimensions using qutrits entangled
in orbital angular momentum.\par

In our experiment the entangled qutrits were produced via type-I
spontaneous parametric down-conversion using a BBO crystal (Beta
Barium Borate) of 1,5 mm thickness pumped by an Argon ion laser
operating at 351nm and having 120 mW of light power. Since we
were not interested in polarization entanglement but in the
entanglement of the orbital angular momentum we chose type-I
down-conversion thus exploiting the greater efficiency of
producing entangled photon pairs. We used the energy-degenerated
case where both entangled photons had a wavelength of 702nm.

Due to their helical  wave fronts the electromagnatic field  of
photons having an orbital angular momentum has a phase
singularity. There the intensity has to vanish resulting in a
doughnut-like intensity distribution. These light fields can be
described by means of Laguerre Gaussian ($LG_{pl}$) modes with
two indices $p$ and $l$. The $p$-index identifies the number of
radial nodes observed in the tranversal plane and the l-index the
number of the 2$\pi$-phase shifts along a closed path around the
beam center. The latter determines the amount of orbital angular
momentum in units of $\hbar$ carried by one photon
\cite{Allen92a,He95a,Friese96a}. In all our experiments we only
considered LG modes with an index $p=0$ whereas the $l$-index of
the entangled photons varied from -2, -1,...to ..1, 2. Since the
$LG_{pl}$ modes form an orthogonal basis they can be used to
realize discrete higher dimensional entangled systems.

A common technique to produce $LG_{0l}$ modes out of the Gaussian
mode is to use computer generated holograms which are usually
transmission gratings with dislocations in the center
\cite{Bazhenov90a,Arlt98a}. Inversely such a hologram in
connection with a single-mode optical fiber can be used to
identify a certain $LG_{0l}$ mode \cite{Mair01a}. It has also
been demonstrated that by displacing such a hologram it is
possible create superposions of the $LG_{00}$ (=Gaussian) and the
corresponding $LG_{0l}$ mode with well-defined amplitudes and
relative phases. In the present experiment in order to identify
the $LG_{01}, LG_{0-1}, LG_{02}$ and $LG_{0-2}$ modes blazed
transmission phase gratings with a period of 20 $\mu \mathrm{m}$
were used which had a diffraction efficiency of  about 85\%.

The experimental demonstration of the concentration was performed
in two steps . Besides our earlier confirmation of conservation
of orbital angular momentum \cite{Mair01a} we also demonstrated
entanglement of the $LG_{0-1}$, $LG_{00}$ and $LG_{01}$ modes by
violating a generalized CHSH type Bell inequality
\cite{Vaziri02b}. It was therefore reasonable to expect that those
results also hold for the extension to $LG_{0-2}$ and $LG_{02}$.
Thus first we confirmed this issue. Restricting ourselves to the
case of a pump beam having no orbital angular momentum, that is an
$LG_{00}$ (Gaussian) mode, it was shown that the entangled
down-conversion state was given by
$C_{00}|00\rangle+C_{11}|11\rangle+C_{22}|22\rangle$, where the
numbers in the kets are equivalent to the absolute value of the
$l$-index $|l|$ of photon 1 and photon 2 respectively. Using the
same techniques as in earlier experiments \cite{Mair01a} the down
converted photons on each side were projected onto the respective
eigenstates via computer generated holograms. The amplitudes
$C_{ij}$ were determined from the coincidence count rates which
are a measure for the probabilities. In order to demonstrate the
entanglement the state was also measured in a rotated basis, i.e.
the down converted photons were projected onto superpositions of
$LG$ modes (Fig.\ref{kurven} upper row). As it is discussed in an
earlier paper\cite{Vaziri02a} this can be achieved by displaced
holograms.

In the second step which we actually demonstrated entanglement
concentration. This we showed by converting the initial entangled
state having non-equal relative amplitudes into a state with
equal amplitudes representing a maximally entangled state.

\begin{figure}
\begin{center}
\includegraphics[width=0.9\textwidth,angle=0]{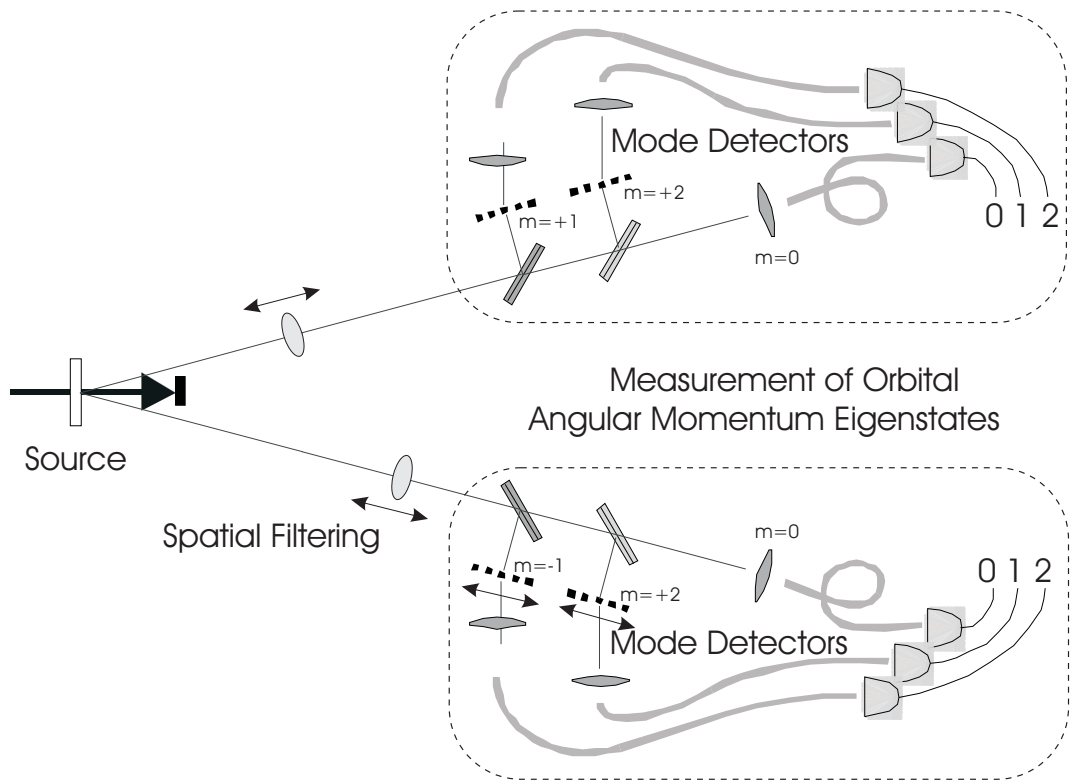}
\end{center}
\caption{Experimental setup for the entanglement concentration.
After type-I parametric down-conversion two lenses are used for
spatial mode filtering by which the relative amplitudes between
different LG modes are changed. The second part of the setup
projects the photons onto the orbital angular momentum eigen
states. This is done by using mode detectors consisting of
computer generated holograms and single mode optical fibers.}
\label{setup}
\end{figure}

The experimental setup is shown in Fig.\ref{setup}. The Gaussian
($LG_{00}$) is focused on the BBO crystal where the entangled
pairs are produced. These are emitted from the crystal at an
angle of $4^{\circ}$ off the pump beam and are coupled into
optical fiber couplers  via a lens on each side.

For convenience the mode identification was done using a
probabilistic method since the count rates were sufficiently high.
However a deterministic mode separator
\cite{Vasnetsov01a,Leach02a} would be possible but significantly
more complicated. Two non-polarizing beam splitters, the first one
having a transmission to reflectivity ratio of 2:1 the second one
of 1:1, redirected each photon with probability of 1/3 to each of
three mode detectors. As mentioned above the combination of a
computer generated hologram and a single mode optical fiber acts
as a mode detector. Using the respective holograms and single
photon detectors we were able to detect $LG_{0-2}, LG_{0-1},
LG_{00}, LG_{01}$ and $LG_{02}$ modes. The coincidence logic was
designed to record the coincidences between an arbitrary pair of
detectors each on one side. This gives 9 possible coincidence
count rates.\par

It has been experimentally observed \cite{Mair01a} that the
emission probability for the higher order entangled modes
decreases with the index $l$. Besides, in order to have a maximum
collection efficiency of the down converted photons it is
important to adapt their beam parameters to the spatial mode
which can be coupled into the single mode fibers
\cite{Kurtsiefer00a}. Usually the proper alignment is achieved by
overlapping the waist of the down converted beam on the crystal
with the waist of an auxiliary beam being sent in a reverse
direction through the single mode fiber and the coupling lenses.
However the waist size of the $LG$ modes grows with their l-index.
Therefore there does not exist a common setting of the coupling
lenses which can couple the $LG_{00}, LG_{01}$ and the $ LG_{02}$
modes all with maximum efficiency. In order to measure the initial
down-converted state emitted by the crystal we had to proceed in
the following way. First the coupling lenses were positioned to
have maximal collection efficiency for the $LG_{00}$ mode. Then
the lens positions were changed to maximize the collection
efficiencies for the $LG_{01}$ and the $LG_{02}$ mode
respectively.

Therefore the measured amplitudes depend on the positioning of the
coupling lenses. Therefore it is possible by varying their
positions to couple in one mode more effectively than the other
one. This method can be considered as a kind of filtering because
part of the photon state emitted in the modes for which the
collection efficiency of the setup ist not optimal is lost.

\begin{figure}
\begin{center}
\includegraphics[width=0.9\textwidth,angle=0]{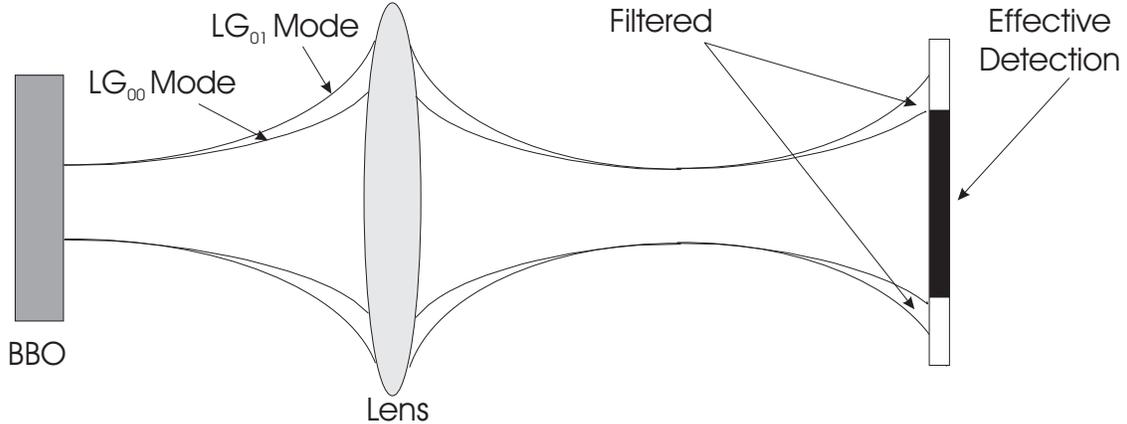}
\end{center}
\caption{The principle of the mode filtering for two different
modes. The two modes $LG_{00}$ and $LG_{0|1|}$ posses the same
beam waist at the crystal. However because of their different
beam divergences they are not detected with the same efficiency.
This effect is exploited to filter the more intense $LG_{00}$
mode. } \label{filter}
\end{figure}

\begin{figure}
\begin{center}
\includegraphics[width=0.7\textwidth,angle=0]{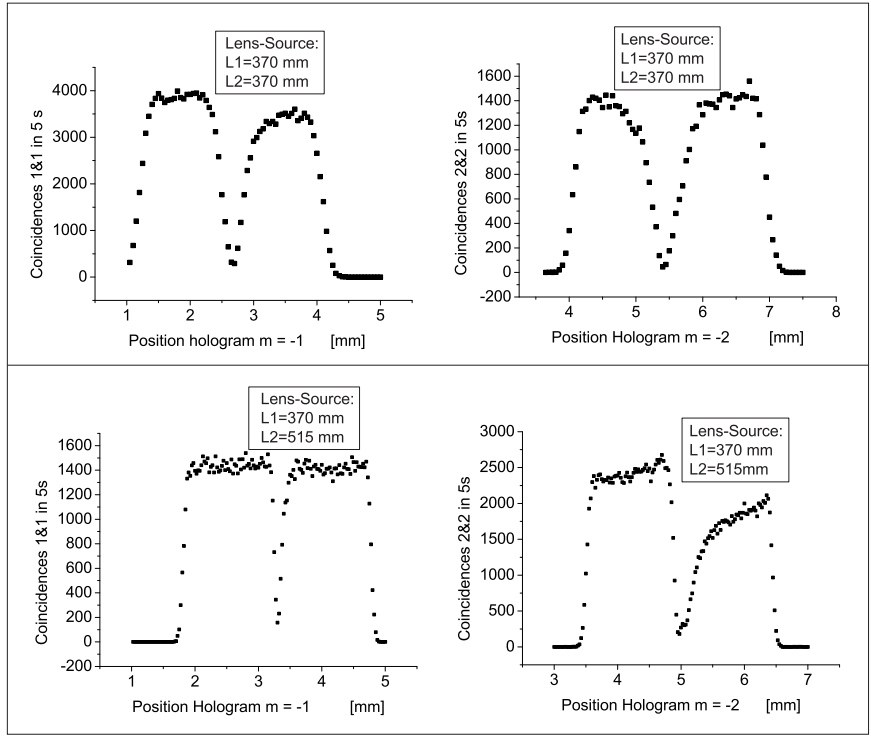}
\end{center}
\caption{Measurement of the entangled states in a superposition
basis. Upper row: before concentration, lower row: after
concentration. In both curves the dip is the signature for
entanglement. The measurement of a mixed state in the
superposition basis would result in equally distributed
coincidences forming curves with no or low contrast.}
\label{kurven}
\end{figure}

In Fig.\ref{filter} the filtering effect is illustrated in one of
the down-conversion arms for two different LG modes (e.g.
$LG_{00}$ and $LG_{0|1|}$).

The refractive index of the mono-mode fibers determines a certain
angle of acceptance for incoming light. The position of the lens
is chosen such that the LG mode with the lower emission
probability (here $LG_{0|1|}$) has a good overlap with the
aceptance mode of the fiber. \footnote{For simplicity the
hologram which would be needed to transform the LG modes with
$|l| \neq 0$ into the Gaussian mode in order to make them
detectable via mono-mode fibers is left out in Fig.\ref{filter}}
As a result the LG mode with the higher emission probabilty (here
$LG_{00}$) has not an optimal overlap with the aceptance mode of
the fiber which causes a filtering of the amplitude of this LG
mode. The same arrangement with another distance can be used in
the other arm of the down-conversion to achieve a filtering of the
$LG_{0|1|}$ mode with respect to the $LG_{0|2|}$ mode.

It is also important to mention that as a consequence of
entanglement each lens acts as a non-local filter on both sides.
The filtering action of a lens on one side projects also the
corresponding modes on the other side. By varying the distance of
the two coupling lenses from the crystal just by the same amount
one would be able to equalize the amplitudes of two modes only.
It is only by choosing asymmetric positions and by exploiting the
entanglement that it is possible to achieve nearly the same
coincidence count rates for all three different modes.

In order to identify the state emitted by the crystal we
proceeded as described above choosing three different lens
positions for collecting the $LG_{00}, LG_{0|1|}$ and $LG_{0|2|}$
respectively each with the same efficiency. Afterwards the
amplitude of each mode was taken from that choice of setup having
the maximum collection efficiency for the corresponding mode. It
is only by this way that the same collection efficiency is ensured
for all modes. Using detectors with nearly equal detection
efficiency the normalized initial state was found to be

\begin{equation}
\psi_{initial}=0,80|00\rangle+ 0,44|11\rangle+0,41|22\rangle
\label{a}
\end{equation}

The amplitudes were calculated from the coincidence count rates
which represent the probabilities for detecting the photon pair
in the corresponding $LG$ mode. For convenience we chose the
basis for describing the initial state such as having no relative
phases between the components. Such a choice is always possible.
By scanning all holograms only horizontally these relative phases
remain unchanged.

In order to demonstrate that the initial state is entangled it
was also measured in bases rotated in Hilbert space. This was done
by displacing the holograms. A displaced
$LG_{01}$($LG_{02}$)-hologram projects an incoming mode onto a
certain superposition of the $LG_{00}$ and the
$LG_{01}$($LG_{02}$) depending on its displacement
\cite{Vaziri02a}. Therefore in one beam the $LG_{01}$- and the
$LG_{02}$-holograms were displaced while in the other beam the
corresponding holograms performed a scan of the mode of the
incoming photons. The resulting coincidences are shown in
Fig.\ref{kurven} upper row.

As shown in an earlier work \cite{Mair01a} the high visibility
($\sim 80 \%$) of these curves in the rotated basis can be viewed
as a signature of entanglement. whereas a mixed state would lead
to a equally distributed coincidence measurement of the LG modes
resulting in curves having low contrast.

\begin{figure}
\begin{center}
\includegraphics[width=0.9\textwidth,angle=0]{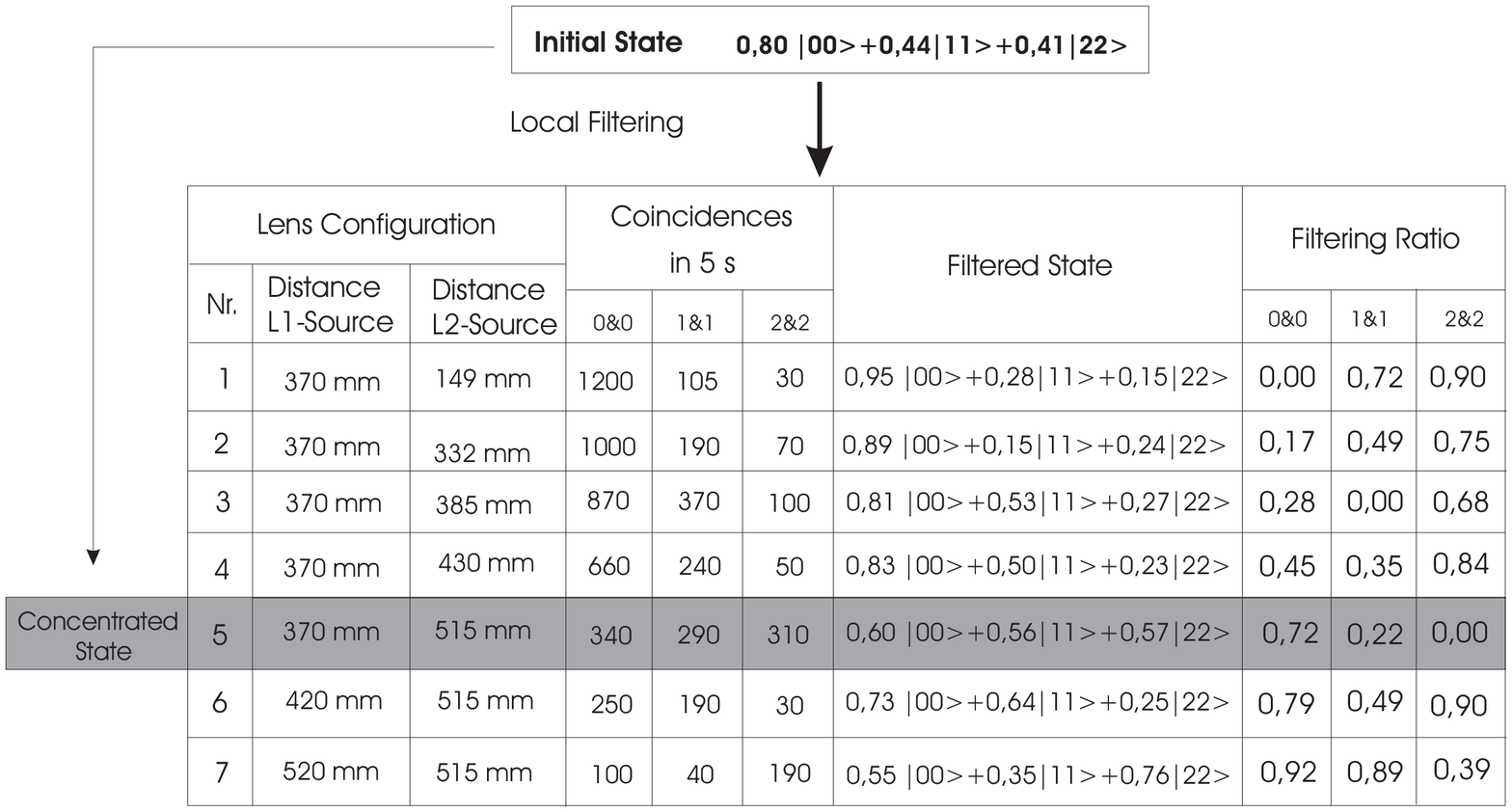}
\end{center}
\caption{"Procrustean" filtering method. Each lens configuration
in the setup cases a specific filtering of the three modes
$LG_{00}, LG_{0|1|}$ and $LG_{0|2|}$. In general the filter
density is different for each mode and depends on the positions
of the two lenses L1 and L2. Specially the lens configuration 5
causes the concentration of the initial state. The concentrated
state has nearly equal amplitudes. Typical errors for the
amplitudes of the filtered states are about 0,02. }
\label{coincidence_tabelle_neu}
\end{figure}

As discussed above different lens positions cause different
filterings of the initial state (1). That means by choosing
various lens positions on both sides different entangled states
can be created out of an initial state. After having identified
the "initial state" this was demonstrated experimentally by for 7
different combinations of lens settings in the two arms and
detecting the coincidences. Out of these the normalized amplitudes
were calculated for the resulting filtered states
(Tab.\ref{coincidence_tabelle_neu}). Each of these lens
configurations can be identified with a certain filter density for
the LG modes. The filter density for each LG mode is defined as
$1-\frac{C_{LC}}{C_{INT}}$ where $ C_{LC}$ and $C_{INT}$ denote
the conincidence count rate at a certain lens configuration and
the coincidence count rates for the initial state respectively.
It is a quantitative measure therefore how a lens configuration
acts as a filter for an LG mode.

The filtered states are not always necessarily more entangled for
any lens configuration, however the filtering action of the lens
configuration Nr. 5 in Tab.2 causes a maximal concentration of the
initial state emitted by the crystal. The concentrated state is
found to be

\begin{equation}
\psi_{concentrated}=0,60|00\rangle+ 0,56|11\rangle+0,57|22\rangle
\label{b}
\end{equation}

which is very close to the three-dimensional maximally entangled
state

\begin{equation}
\psi_{max}=\frac{1}{\sqrt{3}}\left[|00\rangle+|11\rangle+|22\rangle\right]\label{c}
\end{equation}.\par

Since state (1) was the only "initial" state available to us other
lens configurations yielded filtered state which are not
necessarily more entangled. However because of the linearity of
the filtering process there exist possible initial states for
which each of the local filterings in
Tab.\ref{coincidence_tabelle_neu} would cause the concentration
of entanglement. For example given the states $
0,26|00\rangle+0,50|11\rangle+0,83|22\rangle$ or
$0,73|00\rangle+0,63|11\rangle+0,27|22\rangle$ as initial states
the filtering action of the lens configurations Nr. 1 and Nr. 7 in
Tab.\ref{coincidence_tabelle_neu} would cause the concentration
into the maximally entangled state (\ref{c}) respectively.

After we had experimentally equalized the amplitudes of the
initial state (\ref{a}) we wanted to demonstrate that the
resulted filtered state (\ref{b}) was entangled. Therefore we
proceeded in the same way as for demonstrating the entanglement
of the initial state. The state (\ref{b}) was measured in bases
rotated in Hilbert space. Again, the two holograms on one side
were displaced while the two holograms in the other arm made
scans of the modes of the photons.

The measured coincidences are shown in Fig.3 lower row. One can
find that in the $\psi_{concentrated}$ case which is closer to the
maximally entangled state the visbilities are higher. These are
defined as $V=\frac{C_{max}-C_{min}}{C_{max}+C_{min}}$ where $Co$
denotes the coincidence count rates. The corresponding
visibilities in Tab.4 are $86,4\%$for the $LG_{01}$ modes and
$81,5\%$ for the $LG_{02}$ in the $\psi_{initial}$ case and
$94,4\%$ for the $LG_{01}$ modes and $87,3\%$ for the $LG_{02}$
in the $\psi_{concentrated}$ case. The asymmetry in Fig.3 lower
row right is due to an imperfection of the corresponding hologram.
\par
By exploiting simple experimental techniques theses results
clearly show  the possibility to extract maximally entangled
states out of non maximally entangled ones in higher dimensions.
Furthermore the present work demonstrates that using the same
technique it is possible to produce and to identify different
entangled states. The results presented here are not only of
interest for pure states but also for extracting entanglement out
of mixed states. Since, as demonstrated by Horodecki et al.
\cite{Horodecki97a}, all inseparable quantum systems can be
distilled to singlet states by local filtering and the Bennett et
al. distillation protocol \cite{Bennett96b}. Therefore it can be
expected that the ideas and techniques presented here will be of
importance for future quantum communication networks over long
distances which necessarily will need some kind of entanglement
enhancement procedure.

This work was supported by the Austrian FWF, project F1506.

\vspace{-0.2cm}


\vspace{-1cm}


\bibliographystyle{unsrt}


\end{document}